\documentclass[pra,tightenlines,showpacs,nofootinbib]{revtex4}
\usepackage{dcolumn,graphicx}

\begin{document}
\preprint{UNR Mar 2001-\today }
\title{ Correlated many-body treatment of Breit interaction with
application to cesium atomic properties and parity violation}

\author{Andrei Derevianko\email{andrei@unr.edu}}
\affiliation { Department of Physics, University of Nevada, Reno,
Nevada 89557}

\date{\today}

\begin{abstract}
Corrections from Breit interaction to basic properties of
atomic $^{133}$Cs are determined in the framework of third-order
relativistic many-body perturbation theory. The corrections
to energies, hyperfine-structure constants, off-diagonal hyperfine
$6S-7S$ amplitude, and electric-dipole matrix elements are tabulated.
It is demonstrated that the Breit corrections to correlations
are comparable to the Breit corrections at the Dirac-Hartree-Fock level.
Modification of the parity-nonconserving (PNC) $6S-7S$ amplitude due to Breit
interaction is also
evaluated; the resulting weak charge of
$^{133}$Cs shows no
significant deviation from the prediction of the standard model of elementary particles.
The neutron skin correction to the PNC amplitude is also estimated to be
-0.2\% with an error bound of 30\% based on the analysis of recent experiments with
antiprotonic atoms.
The present work supplements publication [A. Derevianko, Phys.\ Rev.\ Lett.\
{\bf 85}, 1618 (2000)] with a discussion of the
formalism and provides additional numerical results and updated discussion
of parity violation.
\end{abstract}


\maketitle

\section{Introduction}

The most accurate to date measurement of parity-nonconservation
(PNC)~\cite{Khr91,BouBou97} in atoms has been carried out by
Wieman and co-workers using
$^{133}$Cs~\cite{WooBenCho97,BenWie99}. The observed weak charge
of the nucleus, $Q_{\rm W}$, is determined as a combination of the
experimental PNC amplitude $E_{\text{PNC}}$ of the
$6S_{1/2}-7S_{1/2}$ transition and a theoretical atomic-structure
parameter $E_{\text{PNC}}/Q_{\rm W}$. Such determined $Q_{\rm W}$
provides powerful constraints on possible extensions to the
standard model (SM) of elementary particles. The achieved
precision in experiments~\cite{WooBenCho97,BenWie99} is 0.35\%;
however the required atomic structure parameter has been
calculated only with an accuracy of about
1\%~\cite{BluJohSap90,DzuFlaSus89}, limiting the accuracy of
determination of the weak charge. Presently it is
understood~\cite{Der00,DzuHarJoh01,KozPorTup01} that a detailed
account of the Breit corrections to basic atomic properties is
required to reach the next level of precision in {\em ab initio}
relativistic calculation of PNC amplitudes. In particular, the
Breit correction to the  $6S_{1/2}-7S_{1/2}$ PNC amplitude in
$^{133}$Cs accounts for a dominant part of the
deviation~\cite{BenWie99} of determined weak charge from the
prediction of the standard
model~\cite{Der00,DzuHarJoh01,KozPorTup01}.

The purpose of this paper is to provide a detailed discussion of
the formalism employed  in Ref.~\cite{Der00a,Der00} and to tabulate
additional numerical results. Since the
publication of Ref.~\cite{Der00a,Der00} several calculations of
the Breit correction to properties of cesium atom have been
carried out~\cite{DzuHarJoh01,Sus01,KozPorTup01} and a comparison
between different approaches is  also presented here. We also
calculate a value for the neutron ``skin'' correction to the
PNC amplitude based on the analysis of experiments with
antiprotonic atoms~\cite{TrzJasLub01}.

The major difference between the present analysis and earlier works
on the Breit interaction in multi-electron
atoms~\cite{Gra61,Gra65,Kim67,SmiJoh67,ManJoh71,LinMarYnn89,JohBluSap88}
is the systematic treatment of correlation effects, i.e. contributions beyond
self-consistent Breit-Coulomb-Hartree-Fock (BCHF)
formulation~\cite{Gra65,ManJoh71,LinMarYnn89}. These correlation
effects are estimated here in the framework of relativistic
many-body perturbation theory. It is demonstrated that these
additional contributions are comparable to the lowest-order BCHF
corrections for almost all considered atomic properties.

The paper is organized as follows.
In Section~\ref{SecMethod} we describe the employed many-body
formalism. Numerical results are tabulated and discussed in Section~\ref{Sec_NumRes}.
We consider Breit corrections
to energies, hyperfine-structure constants, off-diagonal hyperfine
$6S-7S$ amplitude, and electric-dipole matrix elements.
The Breit correction to parity non-conserving $6S-7S$ amplitude is also
evaluated in Section~\ref{Sec_PNC}.

\section{ Method }
\label{SecMethod}%
The Breit interaction~\cite{Bre29,Bre30,Bre32} is a two-particle
interaction caused by an exchange of transverse photons between
atomic electrons. Qualitatively, it describes a magnetic
interaction between electrons (so-called Gaunt interaction) and
retardation effect. Its low-frequency form in the Coulomb gauge,
employed here, is given by\footnote{Unless specified otherwise,
atomic units $\hbar=|e|=m_e=1$ are used throughout the paper.}
\begin{equation}
B=\sum_{i<j}-\frac{1}{2r_{ij}}\left\{
{\mathbf \alpha}_{i}\cdot\mathbf{\alpha }_{j}+\left(
\mathbf{\alpha}_{i}\cdot\widehat{r}_{ij}\right)  \left(
\mathbf{\alpha}_{j}\cdot\widehat{r}_{ij}\right)  \right\} \, ,
\label{Eqn_Breit}
\end{equation}
where $\mathbf{\alpha}$ are Dirac matrices and $r_{ij}$ is a
distance between electrons. The omitted frequency dependence
constitutes as little as 1-2\% of the total Breit correction to the
energies for the atomic ground states\cite{ManJoh71}. The
present goal of {\em ab initio} relativistic calculations in Cs is
to reach an overall accuracy of 0.1\%. Since the Breit corrections to
basic properties of Cs are below 1\%, we ignore the frequency
dependence in the present analysis. It is worth noting that a
consistent inclusion of the frequency dependence in the Breit
interaction would require simultaneous treatment of QED
self-energy correction\cite{LinMarYnn89}.

\subsection{Many-body perturbation theory and Breit interaction}
The many-body Hamiltonian of an atomic system can be generally represented as
\begin{equation}
H=H_{0}+T=\sum_{i}h_{0}\left(  i\right)  +\frac{1}{2}\sum_{ij}t(i,j)\,,
\end{equation}
where
\begin{equation}
h_{0}\left(  i\right)  =c\left(
\mathbf{\alpha}_{i}\cdot\mathbf{p}_{i}\right)  +\beta
c^{2}+V_{\mathrm{nuc}}(i)
\end{equation}
is a one-particle Dirac Hamiltonian for an electron including Coulomb
interaction with the nucleus $V_{\mathrm{nuc}}(i)$ and $t(i,j)$ represents
two-particle interactions. To effectively minimize the perturbing two-particle
interactions one introduces a potential $U(i)$ and rewrites the Hamiltonian as%
\begin{equation}
H=\sum_{i}\left\{  h_{0}\left(  i\right)  +U(i)\right\}  +\left\{  \frac{1}%
{2}\sum_{ij}t(i,j)\,-\sum_{i}U\left(  i\right)  \right\}  .\label{Eq_H}%
\end{equation}
For atoms with one valence electron $v$ outside a closed-shell
core a many-body wavefunction $|\Psi_{v}\rangle$ in the
independent-particle approximation is a Slater determinant
constructed from core and valence single-particle orbitals
$\phi_{i}$. These orbitals  satisfy one-particle
Dirac equation%
\begin{equation}
\left(  h_{0}+U\right)  \phi_{i}=\varepsilon_{i}\phi_{i}\,.
\label{Eqn_OneParticle}
\end{equation}
The potential $U$ is usually chosen to be spherically-symmetric
and label
$i$ is a list of conventional quantum numbers $\left\{  n_{i},j_{i}%
,l_{i},m_{i}\right\}$ for bound states, with $n_i$ replaced by
$\varepsilon_{i}$ for continuum. With the complete set of
single-particle states $\phi_{i},$ the Hamiltonian,
Eq.(\ref{Eq_H}), can be recast into the second-quantized form
\begin{equation}
H=\sum_{i}\varepsilon_{i}\,a_{i}^{\dagger}a_{i}+\sum_{ij}(-U)_{ij}%
a_{i}^{\dagger}a_{j}+\frac{1}{2}\sum_{ijkl}t_{ijkl}\,a_{i}^{\dagger}%
a_{j}^{\dagger}a_{l}a_{k}\,.
\end{equation}
Only certain combinations of positive-- and negative--energy solutions
of the Dirac equation~(\ref{Eqn_OneParticle})
are retained in relativistic many-body Hamiltonian ({\em no-pair}
approximation~\cite{BroRav51}).
The reader is directed to Ref.~\cite{SavDerBer99,SapCheChe99} and references
therein for a detailed discussion of the problem of negative-energy states.

We follow a convention of Ref.~\cite{LinMor86}  and label core
orbitals as $a,b\ldots$, excited (virtual) orbitals as
$m,n,\ldots$, and valence orbitals as $v,w$. Indexes $i,j,k,l$
range over both core and virtual orbitals. In this notation the
lowest-order wavefunction is $|\Psi_{v}\rangle^{\left(  0\right)  }%
=a_{v}^{\dagger}\,|0_{\mathrm{core}}\rangle$, where quasi-vacuum
state $|0_{\mathrm{core}}\rangle=\left(
\prod_{a\in\mathrm{core}}a_{a}^{\dagger }\right)  |0\rangle$
represents a closed-shell atomic core. Introducing normal form of
operator products, $:\cdots:$, defined with respect to
$|0_{\mathrm{core}}\rangle$ one can rewrite a two-particle
operator $T$ as a sum of zero--, one--, and two--body
contributions~\cite{LinMor86}
\begin{eqnarray*}
 T^{(0)} &=& \frac{1}{2} \sum_b t_{bb} \, , \\
 T^{(1)} &=& \sum_{ij} t_{ij} :a_{i}^{\dagger} a_{j}: \, , \\
 T^{(2)} &=& \frac{1}{2}\sum_{ijkl}t_{ijkl}\,
            :a_{i}^{\dagger}a_{j}^{\dagger}a_{l}a_{k}:\, ,
\end{eqnarray*}
with $t_{ij}= \sum_a (t_{iaja}-t_{iaaj})$. In this notation the
Hamiltonian reads
\begin{eqnarray}
H^{\prime}= &  \sum_{i}\varepsilon_{i}\,:a_{i}^{\dagger}a_{i}:+\nonumber\\
&  \sum_{ij}\left\{  t_{ij}-U_{ij}\right\}  :a_{i}^{\dagger}%
a_{j}:+\label{Eq_Hprime}\\
&  \frac{1}{2}\sum_{ijkl}t_{ijkl}\,:a_{i}^{\dagger}a_{j}^{\dagger}a_{l}%
a_{k}:\,.\nonumber
\end{eqnarray}
Zero--body contribution to the total Hamiltonian $H$ has been
discarded since it does not affect the properties of valence
states. It is worth emphasizing that the Breit and Coulomb interactions are
of course two-{\em particle} operators; reference to zero-, one-, and
two-{\em body} parts arises due to the separation into the normal forms of operator
products and is just a matter of convenience.

In the case at hand, the two-particle interaction
\[
    T= C + B
\]
is a sum of the instantaneous Coulomb interaction
$C=\sum_{i<j}\frac{1}{r_{ij}}$ and the Breit interaction $B$,
Eq.~(\ref{Eqn_Breit}). Corresponding two-particle matrix elements
are designated as $c_{ijkl}$ and $b_{ijkl}$. The Coulomb interaction dominates
and we distinguish two possibilities in defining the effective
potential $U$ in Eq.~(\ref{Eqn_OneParticle}): traditional
Coulomb-Hartree-Fock (CHF)  potential $U^{\rm CHF}$ and
Breit-Coulomb-Hartree-Fock (BCHF) potential $U^{\rm BCHF}$, where
the Breit and Coulomb interactions are treated on the same
footing. To differentiate between the two resulting  eigensystems
of Eq.~(\ref{Eqn_OneParticle}) we will add bar to the quantities
pertaining to the BCHF case, e.g. $\bar{\varepsilon}_i, \bar{a}_i,
\bar{a}^\dagger_i$.

The conventional Coulomb-Hartree-Fock (CHF) equation reads
\begin{equation}
   \left( h_{0} + U^{\rm CHF} \right) \phi_i = \varepsilon_i   \phi_i \, ,
\end{equation}
$U^{\rm CHF}$ being mean-field Hartree-Fock potential; this
potential contains direct and exchange Coulomb interactions of
electron $i$ with core electrons. A set of CHF equations is solved
self-consistently for core orbitals; valence wavefunctions and
energies are determined subsequently by ``freezing'' the core
orbitals. The Breit-Coulomb-Hartree-Fock (BCHF) approximation
constitutes introduction of the Breit interaction simultaneously
with the Coulomb interaction into the above CHF equation
\begin{equation}
   \left( h_{0} + U^{\rm BCHF} \right) \bar{\phi}_i =
\bar{\varepsilon_i} \bar{\phi_i} \, . \label{Eqn_BCHF}
\end{equation}
Compared to the CHF equations, energies, wave-functions, and the
Hartree-Fock potential are modified. We discuss a relation between
CHF and BCHF methods and the associated relaxation effect in
Section~\ref{Sec_relax}.

To simplify the second-quantized Hamiltonian,
Eq.~(\ref{Eq_Hprime}), we use the fact that matrix elements of the
Hartree-Fock potentials are $\langle \phi_i | U^{\rm CHF} | \phi_j
\rangle = c_{ij}$ and $\langle \bar{\phi}_i | U^{\rm BCHF} |
\bar{\phi}_j \rangle = \bar{c}_{ij} + \bar{b}_{ij}$. In the
Coulomb-Hartree-Fock case the Hamiltonian reduces to a sum of the
conventional Coulomb Hamiltonian
\begin{equation}
H'_C =  \sum_{i}\varepsilon_{i}\,:a_{i}^{\dagger}a_{i}:+
  \frac{1}{2}\sum_{ijkl}c_{ijkl}\,:a_{i}^{\dagger}a_{j}^{\dagger}a_{l}a_{k}:\,
\label{Eqn_HC}
\end{equation}
and the Breit correction
\begin{equation}
\delta_B H'_C = \sum_{ij} b_{ij} :a_{i}^{\dagger}a_{j}:+
  \frac{1}{2}\sum_{ijkl} b_{ijkl}\,:a_{i}^{\dagger}a_{j}^{\dagger}a_{l}a_{k}:\,.
\label{Eqn_HCdB}
\end{equation}
In the case of equivalent treatment of the Breit and Coulomb
interactions (BCHF case) the corresponding Hamiltonian is less
complicated
\begin{equation}
H'_{C+B} =
\sum_{i}\bar{\varepsilon_{i}}\,:\bar{a}_{i}^{\dagger}\bar{a}_{i}:+
  \frac{1}{2}\sum_{ijkl}( \bar{c}_{ijkl} + \bar{b}_{ijkl})
  \,:\bar{a}_{i}^{\dagger} \bar{a}_{j}^{\dagger} \bar{a}_{l}
  \bar{a}_{k}:\,,
\label{Eqn_HCpB}
\end{equation}
since the effective one-body Breit term in Eq.~(\ref{Eqn_HCdB})
has been ``transformed away'' by a proper choice of one-particle
states.

\begin{figure}
\centerline{\includegraphics*[scale=1.00]{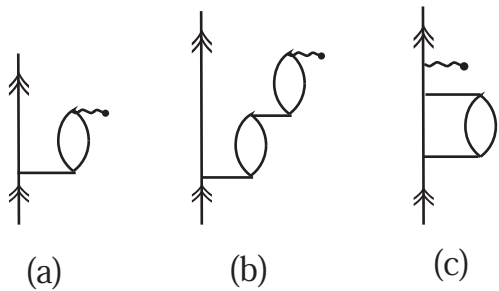}}
\caption{ Sample second-- and third--order Brueckner-Goldstone diagrams
representing many-body contributions to matrix elements. Diagrams (a) and (b)
arise in the random-phase approximation and (c) is the Brueckner-orbital correction.
\label{Fig_ZRPABOdiag} }
\end{figure}

\begin{figure}
\centerline{\includegraphics*[scale=1.00]{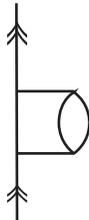}}
\caption{ Representative second-order contribution to an energy of
valence electron.
\label{Fig_EBOdiag} }
\end{figure}

Of course, finding a solution of the Shr\"{o}dinger equation  even
with the traditional many-body Coulomb Hamiltonian,
Eq.~(\ref{Eqn_HC}), is a nontrivial problem. Many-body
perturbation theory~\cite{LinMor86} has proven to be  very
successful in treating contributions beyond the Hartree-Fock
level. In particular, {\em ab initio} relativistic many-body
calculations for alkali-metal atoms have been performed by Notre
Dame and Novosibirsk (Sydney) groups. These and other calculations
have been reviewed recently in Ref.~\cite{Sap98}. An accurate
description of the correlations (i.e. contributions beyond
Hartree-Fock value) plays a crucial role in
high-precision calculations. One of the most striking examples
of the importance of correlations in $^{133}$Cs is
the magnetic-dipole hyperfine-structure (HFS) constant   $A$ of $5D_{5/2}$ level. Here
the Coulomb-Hartree-Fock value, +7.47 MHz, has a sign opposite to
that of experimental value -21.24(5) MHz from Ref.~\cite{YeiSieCer98}.
The dominant correlation corrections to matrix elements
arise because of core-shielding of externally applied fields (e.g.
nuclear fields for HFS constants) and an additional attraction of
a valence electron by an induced dipole moment of the
core~\cite{JohLiuSap96}. The former effect is described by
contributions beginning  at second order (random-phase
approximation (RPA)) and the latter in third order (Brueckner
corrections) of many-body perturbation theory.
Representative many-body diagrams are shown in Fig.~\ref{Fig_ZRPABOdiag}.
Qualitatively, the
Breit correction to a certain Coulomb diagram is proportional to
the value of the Coulomb diagram. Therefore, in addition to lowest-order corrections we
consider the Breit contributions to the dominant RPA and Brueckner
diagrams. It will be demonstrated that these {\em correlated}
Breit corrections in many cases are comparable to the lowest-order
ones. In the BCHF basis the correlated Breit correction to valence energies appears
in the second-order; a sample  diagram is drawn in Fig.~\ref{Fig_EBOdiag}. The
corrections to this class of diagrams is comparable to the modification
at the Hartree-Fock level.

In the present work we employ many-body perturbation theory
(MBPT).
Explicit expressions for contributions to matrix elements
up to the third order were tabulated by \citet{BluGuoJoh87}. These
authors provide formulas for a general perturbing potential with
one-- and two--body parts. In the second and third orders there
are 31 distinct diagrams involving one-body part of the
perturbation and 28 diagrams containing only two-body part.
Certainly calculations of the Breit corrections are less
complicated in the Breit-Coulomb-Hartree-Fock basis, where the
one-body perturbation is absent. Another advantage of the BCHF
basis is an automatic inclusion of important relaxation effect discussed
below. An adequate account for the relaxation effect correction in
the CHF basis would have required fifth-order calculations for matrix elements.

The generalization of MBPT expressions to a simultaneous
treatment of Coulomb and Breit interactions is straightforward:
Coulomb interaction lines are replaced by a sum of Coulomb and
Breit interactions and particle (hole) lines by Breit-Coulomb-Hartree-Fock states
(see Figs.~\ref{Fig_ZRPABOdiag} and~\ref{Fig_EBOdiag}.)
Together with the corrections linear in the Breit interaction
such approach introduces terms nonlinear in the Breit interaction.
Strictly speaking these nonlinear terms have no meaningful theoretical
basis and therefore have to be omitted.
However, the Breit contribution  to atomic properties is relatively small and the
much smaller  terms nonlinear in the Breit interactions can be
neglected at the present level of accuracy.

\subsection{Relaxation effect}
\label{Sec_relax}
\begin{figure}
\centerline{\includegraphics*[scale=0.75]{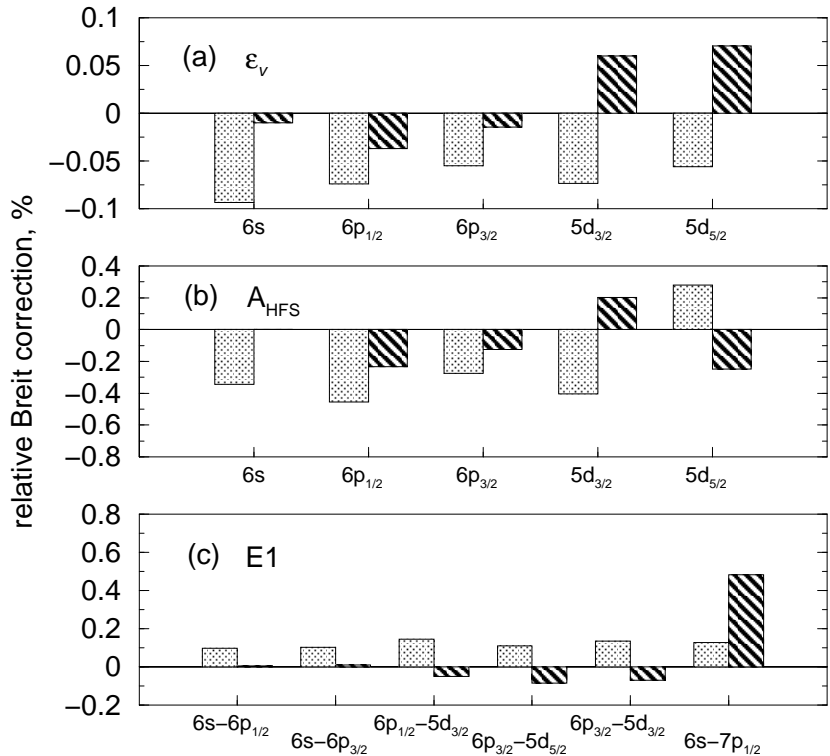}}
\caption{Comparision of Hartree-Fock and first-order relative one-body
Breit corrections to (a)
energies of valence states, (b) hyperfine-structure (HFS) constants  $A$, and (c)
electric-dipole transition amplitudes. The dotted and striped bars
represent first-order  and  Hartree-Fock corrections,
respectively. The relative corrections to the energies and HFS constants are
defined with respect to  experimental values, and electric-dipole
amplitudes with respect to Coulomb-Hartree-Fock values.
\label{Fig_relax} }
\end{figure}

In the CHF basis the first-order corrections to valence energies
$\varepsilon_v$ and matrix elements $Z_{wv}$ due to the one-body
part of the Breit interaction are given by
\begin{eqnarray}
\delta \varepsilon _{v} &=& b_{vv} \, ,
\label{Eq_dB1}\\
 \delta Z_{wv} &= & \sum_{i\not=v} \frac{z_{wi}\,
b_{iv}}{\varepsilon _{v}-\varepsilon
_{i}}+\sum_{i\not=w}\frac{b_{wi}\, z_{iv}}{\varepsilon
_{w}-\varepsilon _{i}}\, . \nonumber
\end{eqnarray}
Similar one-body Breit corrections can be calculated as
differences between lowest-order values found in the
Breit-Coulomb- and Coulomb-Hartree-Fock approximations
\begin{eqnarray}
\delta \varepsilon_{v}^{\rm HF} &=& \bar{\varepsilon}_v - \varepsilon_v \, ,
\label{Eq_dHF}\\
 \delta Z_{wv}^{\rm HF} &=& \langle \bar{\phi}_w| z
| \bar{\phi}_v \rangle - \langle \phi_w| z | \phi_v \rangle.
\nonumber
\end{eqnarray}
In Fig.~\ref{Fig_relax} we present a comparison of the
lowest-order one-body Breit corrections to valence energies,
hyperfine-structure (HFS) constants  $A$ and electric-dipole
transition amplitudes. The dotted and striped bars represent
first-order, Eq.~(\ref{Eq_dB1}), and  Hartree-Fock corrections,
Eq.~(\ref{Eq_dHF}), respectively. There is a striking discrepancy
between the two corrections for all these quantities. For example,
the first-order correction to the $6s$ HFS constant is -0.4\%,
while at the Hartree-Fock level the Breit correction almost
vanishes.

 These large discrepancies are explained by a
``relaxation'' effect, i.e. modification of the Hartree-Fock
potential through adjustment of core orbitals~\cite{LinMarYnn89}.
To illustrate this effect we rewrite the BCHF equation,
Eq.~(\ref{Eqn_BCHF}), as
\begin{equation}\label{Eqn_PertBCHF}
   \left( h_{0} + U^{\rm CHF} + \Delta U  \right) \bar{\phi}_i =
\bar{\varepsilon_i} \bar{\phi_i} \, ,
\end{equation}
where  the perturbing potential is $\Delta U = U^{\rm BCHF} -
U^{\rm CHF}$. Further $ \bar{\phi}_j = \phi_j + \chi_j$, where
$\chi_j$ is a correction to a CFH wavefunction $\phi_j$ due to the
Breit interaction.  In the lowest order these corrections can be
expressed as
\begin{equation}\label{Eqn_Chi}
  \chi_j = \sum'_i \phi_i
  \frac{\langle i | \Delta U | j
  \rangle}{\varepsilon_j-\varepsilon_i}.
\end{equation}
To the first order in the Breit interaction
\begin{eqnarray}
\lefteqn{\Delta U \left(  1\right)     \approx \sum_{a}
\int\phi_{a}^{\dagger}\left(  2\right)  \,b\left(  1,2\right)
\,\phi_{a}\left(  2\right)  d\tau_{2}
+}\nonumber \\
& +\sum_{a}  \int\chi_{a}^{\dagger}\left(  2\right) \,c\left(
1,2\right)  \,\phi_{a}\left(  2\right)
d\tau_{2}   + \label{Eq_dU} \\
& +\sum_{a}  \int\phi_{a}^{\dagger}\left(  2\right) \,c\left(
1,2\right)  \,\chi_{a}\left(  2\right) d\tau_{2} +
\text{exchange}. \nonumber
\end{eqnarray}
Here the first term is an explicit Breit contribution, while in
the second and the third terms the Breit interaction enters
implicitly through corrections to core orbitals. Only the first
term (and its exchange form) are included in the first-order
correction, Eq.~(\ref{Eq_dB1}). As demonstrated by
\citet{LinMarYnn89} for Breit corrections to the energy levels of
Hg, the residual ``relaxation'' terms are large and substantially
modify the first-order corrections. Similar observation has been
made by~\citet{JohBluSap88b} in calculations of Breit corrections
to energies of sodium-like ions. Independent to the present
analysis ( partially published in Ref.~\cite{Der00}) the
relaxation effect in Cs has been recently discussed by
\citet{KozPorTup00}.

At this point it is clear that the inclusion of Breit interaction
in the Coulomb-Hartree-Fock equations (i) greatly simplifies
many-body perturbation expansions, and (ii) automatically accounts
for the significant relaxation effects. In other words, compared
to the traditional Coulomb-Hartree-Fock formulation, the
transformation to the Breit-CHF basis sums many-body diagrams
involving the effective one-body Breit interaction to all orders
of perturbation theory.

\subsection{Construction of Breit-Coulomb Hartree-Fock basis}
\begin{figure}
\centerline{\includegraphics*[scale=0.75]{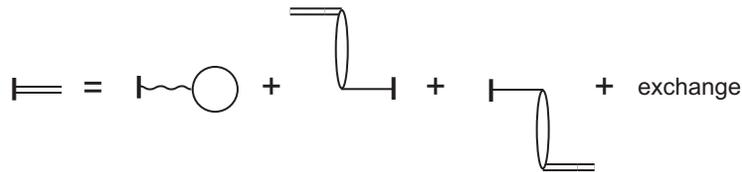}}
\caption{Diagrammatic representation of Eq.~(\protect\ref{Fig_xi}).
Here horizontal solid (wavy) lines represent the Coulomb (Breit) interactions.
Double horizontal lines are expansion coefficients $\xi$ and
little ``stumps'' indicate places where particle or hole lines are to be attached.
\label{Fig_xi} }
\end{figure}

Several methods can be devised for constructing the  Breit-CHF
single-particle basis. For example, one can determine the Breit
corrections to wavefunctions by substituting Eq.~(\ref{Eq_dU})
into Eq.~(\ref{Eqn_Chi}). It is convenient to express the
resulting equations in terms of expansion coefficients $\xi_{ij} =
\langle \phi_i | \chi_j \rangle$
\begin{equation} \label{Eq_BRPA}
\left( \varepsilon _{j}-\varepsilon _{i}\right)
 \,\xi_{ij}=\sum_{a}\tilde{b}_{iaja}+\sum_{ka}\left\{
 \xi_{ka}^{\ast}\tilde{c}_{kiaj}+\xi _{ka}\tilde{c}_{aikj}\right\}
\end{equation}
Here $\tilde{t}_{ijkl}$ is an anti-symmetrized two-particle matrix
element $\tilde{t}_{ijkl}= t_{ijkl}-t_{ijlk}$. Once the
equations~(\ref{Eq_BRPA}) are solved the ``Breit-dressed basis''
can be determined as $ \bar{\phi_j} = \phi_j + \sum'_i \xi_{ij}
\phi_i$. The derived equations are essentially equivalent to the
random-phase approximation or the self-consistent-field method,
with an effective one-body Breit interaction serving as an
external perturbation. The many-body diagrams for the amplitudes
$\xi_{ij}$ are shown in Fig.~\ref{Fig_xi}. By iterating  these
equations one sums a certain class of many-body diagrams to all
orders in the Coulomb interaction.

The resulting equations~(\ref{Eq_BRPA}) are {\em linear} in the
Breit interaction. It is worth noting that the Breit interaction,
Eq.~(\ref{Eqn_Breit}), is an approximation and terms nonlinear in
the Breit interaction have no meaningful theoretical basis.
Therefore linearized equations~(\ref{Eq_BRPA}) are conceptually
more attractive than the self-consistent BCHF method based on an
integration of Eq.~(\ref{Eqn_BCHF}). However, the Breit
contribution  to atomic properties is relatively small and the
much smaller  terms nonlinear in the Breit interactions can be
safely neglected at the present level of accuracy.

An alternative approach to generating the BCHF basis set has
proven to be more numerically robust and was employed in the
present work. Two complete basis sets, CHF $\{ \phi_i \}$ and BCHF
$\{ \bar{\phi}_i \}$ sets, can be related by a unitary
transformation
\[
  \bar{\phi}_j = \sum_i d_{ij} \, \phi_i .\,
\]
Using Eq.~(\ref{Eqn_PertBCHF}) one determines expansion
coefficients $d_{kj} $ and one-particle BCHF energies
$\bar{\varepsilon}_j$ from secular equations
\begin{equation} \label{Eqn_sec}
\left( \varepsilon_k - \bar{\varepsilon}_j \right) d_{kj} + \sum_i
(\Delta U)_{ki} \, d_{ij} = 0\, , \forall  j \,.
\end{equation}
In this work the difference between the two Hartree-Fock
potentials $\Delta U$ was generated using finite-difference
methods. The radial Coulomb-Hartree-Fock basis was approximated
with B-splines~\cite{JohBluSap88} and then transformed into the
radial BCHF basis employing Eq.~(\ref{Eqn_sec}).
Negative-energy states, $\varepsilon_i< -m_e c^2$, were included
in the diagonalization procedure.

To summarize, third-order many-body calculations were performed
in the Breit-Coulomb-Hartree-Fock basis
with the two-body Breit interaction $B^{(2)}$ treated on equal footing with the residual
Coulomb interaction.
Sample many-body diagrams are presented in
Fig.~\ref{Fig_EBOdiag} and \ref{Fig_ZRPABOdiag}. Contributions of negative-energy
states, discussed for example in Ref.~\cite{SavDerBer99}, were
also included and found to be relatively small~\cite{Der00a}. Two
series of third-order calculations were performed, first with the
Breit and Coulomb interactions fully included using the Breit-CHF basis
set, and second in the CHF basis set without the Breit interaction
and negative-energy states. The obtained differences are the Breit
corrections analyzed in the following sections.

Numerical calculations were performed using B-spline basis sets
generated in a cavity of radius 75 a.u. This cavity size has been chosen
for numerical consistency with the previous determination
of parity-nonconserving amplitudes by \citet{BluJohSap90}.  The numerical
quasi-spectrum was represented by 100 negative- and 100
positive-energy states for each angular quantum number $\kappa$.
The intermediate-state summations were performed over 75
lowest-energy positive-energy states and 75 highest-energy
negative-energy states for each partial wave $s_{1/2}-h_{11/2}$.

\section{Atomic Properties}
\label{Sec_NumRes}
\subsection{Energies of valence states.
\label{Sec_En}}
\begin{figure}
\centerline{\includegraphics*[scale=0.75]{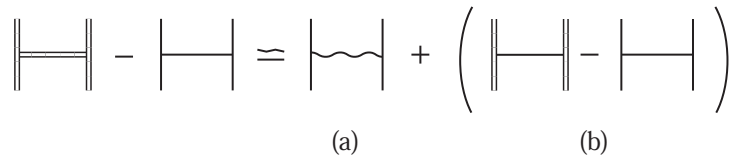}}
\caption{ Separation of Breit correction into (a) two--body
and  (b) one--body (basis shift)  parts.
Here horizontal  solid and wavy lines represent
Coulomb and Breit interactions respectively; double horizontal
line is a sum of Coulomb and Breit interactions. Vertical solid lines
correspond to Coulomb-Hartree-Fock states and vertical double lines
to Breit-CHF states.
   \label{Fig_classes} }
\end{figure}

At the Hartree-Fock level, the Breit interaction contributes less
than 0.1\% to all the energy levels considered in
Fig.~\ref{Fig_relax}. Numerical values for Breit correction at the
Hartree-Fock level are given in Table~\ref{Tab_EBreit}; these
were obtained as differences between
one-particle energies in Breit-CHF and CHF approximations, i.e.
$\bar{\varepsilon}_v - \varepsilon_v$.
 As in the traditional CHF calculations, the
first-order many-body contributions to valence energies  vanish
identically in the BCHF basis. In the second order the corrections
arise due to self-energy diagrams. For each valence state
we perform two calculations with and without the Breit interaction and
take a difference between the two values. Further,
we distinguish between two classes of Breit modifications,
one-body and two-body corrections, as illustrated for a diagram
in Fig.~\ref{Fig_classes}. The one-body contribution arises from a
transformation of the one-particle basis from Coulomb-Hartree-Fock to Breit-CHF.

The calculated values and breakdown on various contributions
are presented in Table~\ref{Tab_EBreit}.
Apparently the Breit correction to the {\em correlation}
part of the energy, $\delta E_{\rm BO}$, is equally important
as the modification in the lowest order, $\delta E_{\rm HF}$.
The interplay of Breit corrections to various many-body diagrams
for the energy of the $7S_{1/2}$ state is remarkable. Here the
two terms  $\delta E_{\rm HF}$ and  $\delta E_{\rm BO}, B^{(1)}$
are almost equal and have opposite signs, the resulting modification
being determined by relatively smaller two-body Breit
correction  $\delta E_{\rm BO}, B^{(2)}$.  From Table~\ref{Tab_EBreit} we see that
generally the two-body Breit  contributions
are smaller than the one-body corrections; the $B^{(2)}$ corrections become
important when the cancellations are involved.

\begin{table}
\centering \caption{Contributions of the Breit interaction to
energies of valence electrons in  cm$^{-1}$.
$E_{\rm CHF}$ are the energies in the Coulomb-Hartree-Fock approximation.
$\delta E_{\rm HF}$ column lists
corrections at the Hartree-Fock level defined as
$\bar{\varepsilon}_v - \varepsilon_v$. Columns $\delta E_{\rm BO},
B^{(1)}$ and $\delta E_{\rm BO}, B^{(2)}$ are the contributions in
the second order due to one-body and two-body Breit interactions
respectively.
 \label{Tab_EBreit}}
\begin{ruledtabular}
\begin{tabular}{lrrddd}
\multicolumn{1}{c }{State}
& \multicolumn{1}{c }{$E_{\rm CHF}$} %
& \multicolumn{1}{c }{$\delta E_{\rm HF}$} %
& \multicolumn{1}{c }{$\delta E_{\rm BO}, B^{(1)}$ } %
& \multicolumn{1}{c }{$\delta E_{\rm BO}, B^{(2)}$ }%
& \multicolumn{1}{c }{Total} \\
\colrule %
$6S_{1/2}$ & -27954&  3.2  &  -4.98  &-0.83  & -2.6   \\
$7S_{1/2}$ & -12112&  1.1  &  -1.1   &-0.28  & -0.26  \\
$6P_{1/2}$ & -18791&  7.5  &  -0.08  &-0.28  &  7.1   \\
$7P_{1/2}$ &  -9223&  2.7  &  -0.1   &-0.10  &  2.5   \\
$6P_{3/2}$ & -18389&  2.9  &  -1.8   &-0.25  &  0.84  \\
$7P_{3/2}$ &  -9079&  1.0  &  -0.56  &-0.09  &  0.38  \\
$5D_{3/2}$ & -14138& -10.2  & -12    &-0.35  & -22    \\
$5D_{5/2}$ & -14163& -11.8  & -14    &-0.33  & -26    \\
\end{tabular}
\end{ruledtabular}
\end{table}

The first study of correlated Breit corrections to the energies of Cs and other
alkali-metal atoms has been performed in Ref.~\cite{SafDerJoh98,SafJohDer99}.
Based on formalism developed in Ref.~\cite{DerJohFri98},
the corrections have been determined as an expectation value
of the Breit correction to the Coulomb-Hartree-Fock many-body Hamiltonian, Eq.~(\ref{Eqn_HCdB}),
\begin{equation}
\delta E_v = \langle \Psi_v^{SD} | \delta_B H'_C | \Psi_v^{SD} \rangle \, .
\end{equation}
Here $| \Psi_v^{SD} \rangle$ is the linearized coupled-cluster
wavefunction limited to single and double excitations from
the reference Slater determinant $a^\dagger_v |0_{\rm core} \rangle$
(CCSD method). The correlations are built
into these wavefunctions. The CCSD formalism accounts for a complete
third order of MBPT with certain classes of diagrams summed to all orders.
However, the random-phase-approximation
sequence of diagrams, important for the self-consistent treatment
of the Breit interaction, is missed starting from the fourth order.
The approach employed by \citet{KozPorTup01} is similar to the present method,
but in Ref.~\cite{KozPorTup01} the full Breit interaction has been
approximated by the Gaunt term and contribution
due to effective two-body interaction $B^{(2)}$ has been neglected.

Comparison of our results with the CCSD values and results of Ref.~\cite{KozPorTup01}
is presented in Table~\ref{Tab_EBreitComp}.
The best agreement is for $P_{1/2}$ states where there is no cancellation
between corrections $\delta E_{\rm HF}$ and $\delta E_{\rm BO}, B^{(1)}$.
The values differ significantly for $S_{1/2}$ states where strong cancellations,
emphasizing higher orders of MBPT,
are present. One should keep in mind that these discrepancies arise
only when the Breit corrections, due to cancellation effects, are small and
do not have an enhanced effect on evaluation of parity non-conserving amplitudes, discussed
in Section~\ref{Sec_PNC}.
Calculations~\cite{KozPorTup01} are less consistent with the present
and CCSD results, with discrepancies caused by approximation of full Breit interaction
by Gaunt term and neglect of two-body part of the Breit interaction.

\begin{table}
\centering \caption{Comparison of Breit corrections to
energies of valence electrons in  cm$^{-1}$.
 \label{Tab_EBreitComp}}
\begin{ruledtabular}
\begin{tabular}{lddd}
\multicolumn{1}{c }{State}
& \multicolumn{1}{c }{This work} %
& \multicolumn{1}{c }{CCSD~\protect\cite{SafJohDer99}} %
& \multicolumn{1}{c }{Gaunt, no $B^{(2)}$~\protect\cite{KozPorTup01}} \\
\colrule %
$6S_{1/2}$ & -2.6  & -1.1  & -4 \\
$7S_{1/2}$ & -0.26 &  0.72 &  0 \\
$6P_{1/2}$ &  7.1  &  6.9  &  9 \\
$7P_{1/2}$ &  2.5  &  2.6  &  2 \\
$6P_{3/2}$ &  0.84 &  0.29 &  2 \\
$7P_{3/2}$ &  0.38 &  0.45 &  0 \\
\end{tabular}
\end{ruledtabular}
\end{table}

\subsection{Magnetic-dipole hyperfine structure constants}
To reiterate discussion in Section~\ref{SecMethod}, we performed
two series of computations: (i) traditional Coulomb and (ii) fully
including Breit interaction. The difference between resulting values
defines the Breit correction. From parametrical argument it is assumed that the leading Breit
corrections arise from induced modifications of the dominant traditional
Coulomb diagrams. Therefore the calculations were limited to the
random-phase-approximation (RPA) and Brueckner diagrams (see Fig.~\ref{Fig_ZRPABOdiag}).
Further the RPA sequence was truncated at the third order.

In calculations of Breit corrections to $^{133}$Cs
hy\-per\-fine-struc\-ture mag\-ne\-tic-di\-po\-le (HFS) constants $A$
nucleus was modelled by a
uniformly magnetized ball of radius $R_m= 5.6748$ fm. The
gyromagnetic ratio for $^{133}$Cs nucleus is
$g_I=0.73789$~\cite{Rag89}.

The breakdown of Breit corrections to various classes of many-body
diagrams is given in Table~\ref{Tab_Acorr}. Clearly the Breit correction
to correlations ($\delta A_{\rm RPA}^{\rm II + III}$, $\delta A_{\rm BO}^{\rm III}$)
is equally important as the modifications in the lowest order $\delta A_{\rm HF}^{\rm I}$.
As an extreme case, almost entire Breit correction to HFS constants of 6S and 7S
states comes from correlations. There is a cancellation of various contributions
to the HFS constants of $P_{1/2}$ and $P_{3/2}$ states. For these states
the contribution of higher-order diagrams not included in the present third-order analysis
can become enhanced. At the same the total Breit corrections to $A_{6S}$
and $A_{7S}$ are expected to be insensitive to higher-order contributions.

\begin{table}
\centering \caption{Breit corrections to
magnetic-dipole hyperfine structure constants $A$
of $^{133}$Cs in MHz. Column CHF lists Coulomb-Hartree-Fock
values. Breit corrections to a class X of many-body diagrams of order $N$
are designated as $\delta A^{N}_{\rm X}$.
Total Breit correction $\delta A_{\rm Total}$ is a sum of
modifications $\delta A^{N}_{\rm X}$.
\label{Tab_Acorr}}
\begin{ruledtabular}
\begin{tabular}{lrdddd}
\multicolumn{1}{c }{State}               &
\multicolumn{1}{c }{CHF}  %
& \multicolumn{1}{c }{$\delta A_{\rm HF}^{\rm I}$ } %
& \multicolumn{1}{c }{$\delta A_{\rm RPA}^{\rm II + III}$ } %
& \multicolumn{1}{c }{$\delta A_{\rm BO}^{\rm III}$ }%
& \multicolumn{1}{c }{$\delta A_{\rm Total}$} \\
\colrule %
$6S_{1/2}$ & 1425.3 & 0.011 &  4.1    &  0.79  &  4.87  \\
$7S_{1/2}$ & 391.6 & -0.029 &  1.1    &  0.08  &  1.15  \\
$6P_{1/2}$ & 160.9 & -0.68  &  0.39   & -0.24  & -0.52  \\
$7P_{1/2}$ & 57.62 & -0.23  &  0.14   & -0.059 & -0.15  \\
$6P_{3/2}$ & 23.92 & -0.06  &  0.10   & -0.008 &  0.034  \\
$7P_{3/2}$ & 8.642 & -0.022 &  0.038  & -0.0020&  0.014 \\
$5D_{3/2}$ & 18.23 &  0.099 &  0.10   &  0.11  &  0.31  \\
\end{tabular}
\end{ruledtabular}
\end{table}

A comparison of our results, partially published in Ref.~\cite{Der00},
with other calculations is presented
in Table~\ref{Tab_AComp}.
The correction to hyperfine constants is very sensitive to
correlations: e.g.,  Ref.~\cite{BluJohSap91} found a numerically
insignificant modification for $A_{6S}$, while
Ref.~\cite{SafJohDer99,Der00a} determined the modification to be
large (-4.64 MHz), and the approach reported here yields +4.87
MHz. In the calculation of Ref.~\cite{BluJohSap91} the correction
was determined as a difference of the Breit-CHF and CHF values, however
such approach misses two-body Breit corrections of comparable
size. In Ref.~\cite{SafJohDer99,Der00a} a second order
perturbation analysis was used for the Breit interaction, but the
important relaxation effect discussed earlier was omitted. The
present calculation incorporates all mentioned diagrams and is
also extended to third order. Motivated by
strong dependence of results~\cite{BluJohSap91,SafJohDer99,Der00} on
many-body corrections,
 \citet{Sus01} derived an analytical
expression for Breit correction to HFS constants of $S$ states.
His results for $6S$ and $7S$ states are in an excellent agreement
with the present calculations. \citet{KozPorTup01} used an approach
similar to Ref.~\cite{Der00}. There is a
cancellation of various contributions for the $P$ states; both higher-order
diagrams and an approximation of the full Breit interaction by the Gaunt term in
Ref.~\cite{KozPorTup01} are the sources of discrepancies between our results
for $6P_{1/2}$ and $7P_{1/2}$ states.

\begin{table}
\centering \caption{Comparison of contributions of Breit
interaction to magnetic-dipole hyperfine structure constants $A$
of $^{133}$Cs in MHz. \label{Tab_AComp}}
\begin{ruledtabular}
\begin{tabular}{ldddd}
               &
\multicolumn{1}{c }{$6S_{1/2}$}  & \multicolumn{1}{c }{$7S_{1/2}$}
& \multicolumn{1}{c }{$6P_{1/2}$} &
\multicolumn{1}{c }{$7P_{1/2}$}        \\
\colrule
 This work\footnotemark[1]          & 4.87 & 1.15 & -0.52 & -0.15  \\
 \citet{KozPorTup01}\footnotemark[2] & 5.0  & 0.8  & -0.2  &  0.0 \\
 \citet{Sus01}\footnotemark[3]       & 4.6  & 1.09 \\ \hline
 \citet{SafJohDer99}\footnotemark[4] & -4.64& -0.83& -0.87 & -0.29\\
 \citet{BluJohSap91}\footnotemark[5] & 0.00 & -0.05& -1.25 &
 -0.39\\
\end{tabular}
\end{ruledtabular}
\footnotetext[1]{ Third-order calculations in the BCHF basis,
Ref.~\cite{Der00}. } %
\footnotetext[2]{ Full Breit interaction is approximated by the Gaunt
term.} %
\footnotetext[3]{ Analytical $\alpha Z$ expansion with $Z=55$.}
\footnotetext[4]{ Second-order calculations in the CHF basis. See
Ref.~\cite{Der00a} for details.} %
\footnotetext[5]{ RPA sequence of diagrams in one-body Breit
interaction}
\end{table}

Hyperfine structure constants sample atomic wavefunctions close to the nucleus
and provide a unique way of testing atomic-structure calculations
of parity-nonconserving amplitudes.
In Table~\ref{Tab_AExpComp} we combine the Breit corrections with the results of
{\em ab initio} all-order Coulomb-correlated calculations~\cite{BluJohSap91}
and compare the results with experimental values. It is clear that the Breit
corrections uniformly improve the agreement. In particular, the theoretical
HFS constants are improved to 0.1\% for $6S_{1/2}$, $7S_{1/2}$, and $7P_{1/2}$ states
except for $6P_{1/2}$ where the discrepancy becomes 0.5\%. While the achieved agreement
in Table~\ref{Tab_AExpComp} is encouraging,
one should keep in mind omitted QED corrections and higher-order contributions
in Coulomb interaction. For example, an estimate~\cite{KarKle52,ArtBeiPlu99}
for {\em hydrogen}-like Cs ion results in a QED correction to HFS constant $A$
of $S$-states at
a few 0.1\%. Due to electron-electron interaction in {\em atomic} Cs
QED corrections can be significantly modified. Correlated
calculations of QED corrections would be beneficial for reaching 0.1\% level
of accuracy needed for interpretation of parity-nonconservation and also
for understanding the role of high-order Coulomb diagrams at 0.1\% precision.

\begin{table}
\centering \caption{Comparison of theoretical and experimental
hyperfine constants $A$ of $^{133}$Cs.
All-order Coulomb-correlated values by \citet{BluJohSap91} are
supplemented with Breit corrections. Deviation from experimental
values are placed in square brackets.
 \label{Tab_AExpComp}}
\begin{ruledtabular}
\begin{tabular}{ldddd}
               &
\multicolumn{1}{c }{$6S_{1/2}$}
& \multicolumn{1}{c }{$7S_{1/2}$}
& \multicolumn{1}{c }{$6P_{1/2}$} &
\multicolumn{1}{c }{$7P_{1/2}$}        \\
\colrule
 Coulomb\cite{BluJohSap91}
               & 2291.00[-0.3\%]&544.09[-0.3\%]&293.92[0.7\%]&94.60   [0.3\%] \\
 Breit         &  4.87          & 1.15         &-0.52       &-0.15     \\
 Total         & 2295.87[-0.1\%]&545.24[-0.1\%]&293.40[0.5\%]&94.45   [0.1\%] \\
 Experiment    & 2298.16        &545.90        &291.93(2)    &94.35(4) \\
\end{tabular}
\end{ruledtabular}
\end{table}

\subsection{Off-diagonal $6S-7S$ hyperfine-structure matrix element}
\label{Sec_OffDiag}
Experiments~\cite{WooBenCho97} on parity-nonconservation (PNC) in
$^{133}$Cs determine the ratio of $6S-7S$ PNC amplitude
$E_{\rm PNC}$ to vector transition polarizability $\beta$. The value of
$\beta$ is difficult to calculate reliably since it vanishes in
the nonrelativistic limit. Following suggestion~\cite{BouBou74},
\citet{BenWie99} determined a supporting ratio of $\beta$ to
off-diagonal magnetic-dipole matrix element $M_{\rm hf}$ with a
precision of 0.16\%. Such an approach eliminates $\beta$ from the
analysis, but requires an accurate value for $M_{\rm hf}$.

The quantity $M_{\rm hf}$ can be expressed in terms of the off-diagonal magnetic-dipole
hyperfine structure constant $A_{6S-7S}$. This constant can be well approximated by
a semiempirical geometric-mean formula~\cite{Hof82}
\begin{equation}
 A_{6S-7S}^{\rm s.e.} \approx \sqrt{A_{6S} A_{7S}} \, ,
\label{Eqn_Aoff}
\end{equation}
where $A_{6S}$ and  $A_{7S}$ are precise experimental hyperfine structure constants.
The accuracy of this expression was investigated in Ref.~\cite{BouPik88,DerSafJoh99,DzuFla00}.
Most recently, \citet{DzuFla00} employed several many-body techniques of increasing
accuracy in the Coulomb
interaction between electrons and found
that this geometric-mean formula is accurate to a fraction of $10^{-3}$.
Here we extend their analysis and rigorously
consider the additional effect of the correlated Breit interaction.
The Breit correction to hyperfine-structure constants $A_{6S}$ and $A_{7S}$ is in the order
of 0.2\% and it can affect the sub-0.1\% accuracy of the Coulomb analysis~\cite{DzuFla00}.

Breit corrections are relatively small.
If  Eq.~(\ref{Eqn_Aoff}) holds, the following relation between the Breit corrections
($\delta A$) has to be satisfied
\begin{equation}
 \frac{ \delta A_{6S-7S}^{\rm s.e.}  }{A_{6S-7S}^{\rm s.e.} }  \approx \frac{1}{2}
 \left\{
 \frac{ \delta A_{6S} }{ A_{6S} } + \frac{\delta A_{7S} }{ A_{7S} } \,
 \right\}
 \, .
\end{equation}
As a result of the correlated Breit calculations we find $\delta A_{6S-7S}=2.4$ MHz.
With $A_{6S-7S}=1120.1$ MHz, the
ratio $\frac{ \delta A_{6S-7S} }{A_{6S-7S}} = 2.1 \times 10^{-3}$.
Using the Breit corrections to HFS constants  $A_{6S}$ and $A_{7S}$ from Table~\ref{Tab_Acorr},
the semiempirical r.h.s. of the above equation is also $2.1 \times 10^{-3}$. Clearly the
accuracy of the geometric-mean formula~(\ref{Eqn_Aoff}) is not affected by the Breit correction.
Qualitatively this can be explained by a close proportionality of $6S$ and $7S$
wavefunctions in the vicinity of the nucleus, where the main contribution to the HFS
constants of $S$ states is accumulated.

\subsection{Electric-dipole transition amplitudes}

Calculated Breit corrections to reduced electric-dipole matrix elements
of various transitions in Cs are presented in Table~\ref{Tab_Dcorr}.
We note that the Breit corrections
to the random-phase approximation diagrams are small compared to the lowest-order and
Brueckner-orbital corrections.
Generally the total corrections are rather small ($\sim 0.1\%$),
with an exception of $6S_{1/2}-7P_{1/2}$ electric-dipole
matrix element. Using the {\em ab initio} all-order Coulomb-correlated value by
\citet{BluJohSap91},
$\langle 6S_{1/2}|| D || 7P_{1/2} \rangle = 0.279$, and adding the Breit correction of 0.0019, one
finds  $\langle 6S_{1/2}|| D || 7P_{1/2} \rangle = 0.281$ in  a better agreement with the
0.284(2) experimental value of ~\citet{ShaMonKhl79}.
The relatively large Breit correction is caused both by an
accidentally small matrix element and by
admixture into $\langle 6S_{1/2} | D | 7P_{1/2}  \rangle$ from a 30 times larger
$7S_{1/2}-7P_{1/2}$ matrix element.

\begin{table}
\centering \caption{Breit corrections to
reduced electric-dipole matrix elements for transitions between low-lying
valence states of Cs atom. Column $\delta D, \%$ represents a ratio (in \%) of the total Breit
correction to a matrix element calculated in a third order of MBPT.
See caption of Table~\protect\ref{Tab_Acorr} for description
of other columns.
\label{Tab_Dcorr}}
\begin{ruledtabular}
\begin{tabular}{llddddd}
\multicolumn{1}{c }{Transition}               &
\multicolumn{1}{c }{CHF}  %
& \multicolumn{1}{c }{$\delta D_{\rm HF}^{\rm I}$ } %
& \multicolumn{1}{c }{$\delta D_{\rm RPA}^{\rm II+III}$ } %
& \multicolumn{1}{c }{$\delta D_{\rm BO}^{\rm III}$ }%
& \multicolumn{1}{c }{$\delta D_{\rm Total}$}
& \multicolumn{1}{c }{$\delta D, \%$} \\
\colrule %
$6S_{1/2}-6P_{1/2}$& 5.278  & 0.00035 & -0.00022 & -0.0011  &-0.00097& -0.02  \\
$6S_{1/2}-6P_{3/2}$& 7.426  & 0.00078 & -0.00045 & -0.0014  &-0.0011 & -0.02 \\[3pt]
$6S_{1/2}-7P_{1/2}$& 0.3717 & 0.0018  & -0.00013 &  0.00021 & 0.0019 &  0.5 \\
$6S_{1/2}-7P_{3/2}$& 0.6947 & 0.00059 & -0.00020 &  0.00011 & 0.00049&  0.07 \\[3pt]
$7S_{1/2}-6P_{1/2}$& 4.413  & 0.0046  & -0.000067&  0.00038 & 0.0049 &  0.1 \\
$7S_{1/2}-6P_{3/2}$& 6.671  & 0.0019  & -0.000011& -0.00030 & 0.0016 &  0.02 \\[3pt]
$7S_{1/2}-7P_{1/2}$& 11.01  &-0.0011  & -0.000050& -0.0018  &-0.0029 & -0.03 \\
$7S_{1/2}-7P_{3/2}$& 15.34  & 0.0007  & -0.00013 & -0.0019  &-0.0013 & -0.009 \\[3pt]
$5D_{3/2}-6P_{1/2}$& 8.978  &-0.0044  & -0.00035 & -0.0082  &-0.013  & -0.2  \\
$5D_{3/2}-6P_{3/2}$& 4.062  &-0.0028  & -0.00019 & -0.0035  &-0.0065 & -0.2  \\
\end{tabular}
\end{ruledtabular}
\end{table}

\section{Parity-nonconserving amplitude $6S_{1/2} \rightarrow
7S_{1/2}$}
\label{Sec_PNC}

The parity-nonconserving  amplitude for the
$6S_{1/2} \rightarrow 7S_{1/2}$ transition in $^{133}\mathrm{Cs}$ can be represented as a sum
over intermediate states $mP_{1/2}$
\begin{eqnarray}
\lefteqn{E_\mathrm{ PNC} = \sum_{m}
\frac{\langle 7S|D|mP_{1/2}\rangle  \langle mP_{1/2} |H_{\rm W}|6S\rangle
}{E_{6S}-E_{mP_{1/2}}}  } \nonumber \\  &+ &
\sum_{m}
\frac{\langle 7S|H_{\rm W}|mP_{1/2}\rangle  \langle mP_{1/2} |D|6S\rangle
}{E_{7S}-E_{mP_{1/2}}}
\, .
\label{Eqn_E_PNC}
\end{eqnarray}
Here $D$  and $H_{\rm W}$ are electric-dipole amplitudes and weak interaction matrix elements,
and $E_{i}$ are atomic energy levels.
The PNC amplitude is expressed  in units of $10^{-11} i |e| a_0 (-Q_{\rm W}/N)$,
where  $N=78$ is the number of neutrons in the nucleus of $^{133}$Cs and
$Q_{\rm W}$ is the weak charge.
In these units the results of past calculations
for  $^{133}\mathrm{Cs}$ are
$E_\mathrm{PNC} = -0.905$, Ref.~\cite{BluJohSap90}, and
$E_\mathrm{PNC} = -0.908$, Ref.~\cite{DzuFlaSus89}.
The former value includes a partial Breit contribution $+0.002$,
and the latter includes none.
The reference many-body Coulomb-correlated amplitude
\begin{equation}
 E^{C}_\mathrm{PNC} = -0.9075   
\label{Eqn_refVal}
\end{equation}
is determined as an average,
with the partial Breit contribution removed from the value of Ref.~\cite{BluJohSap90}.
The major difference between present and previous calculation~\cite{BluJohSap90}
of Breit correction to the PNC amplitude is an additional incorporation
of effective two-body part of the Breit interaction and Breit corrections to
the correlations.

It is convenient to break the total Breit correction $\delta E_\mathrm{PNC}$ into
three distinct parts due to corrections in the weak interaction and dipole
matrix elements, and energy denominators, respectively
\begin{equation}
  \delta E_\mathrm{PNC}^{B}  = E_\mathrm{PNC} (\delta H_{\mathrm W}) + E_\mathrm{PNC} (\delta D) +
E_\mathrm{PNC} (\delta E) \, .
\label{Eq_PNC_corrections}
\end{equation}
For example, the modification of the PNC amplitude due to the Breit corrections to
energies $\delta E_{nS}, \delta E_{mP_{1/2}} $ can be expressed as
\begin{eqnarray}
E_\mathrm{PNC} (\delta E)&=&
- \sum_{m}
\frac{\langle 7S|D|mP_{1/2}\rangle  \langle mP_{1/2} |H_{W}|6S\rangle
}{ (E_{6S}-E_{mP_{1/2}})^2 }  \left( \delta E_{6S} - \delta E_{mP_{1/2}} \right)
\nonumber \\  &+& \mathrm{c.c.}(7S\leftrightarrow 6S) \, ,
\end{eqnarray}
where the last term stands for the complex conjugate of the first term
with  $6S$ and $7S$ states interchanged. The Breit corrections to energies and
dipole matrix elements were discussed in the preceding sections;
here we focus on corrections to the weak matrix elements.

The overwhelming contribution from parity-violating interactions arises from
the Hamiltonian
\begin{equation}
 H_{\rm W} = \frac{G_F}{\sqrt{8}} Q_{\rm W} \rho_\mathrm{nuc}(r) \gamma_5 \, ,
\label{Eqn_Hw}
\end{equation}
where $G_F$ is the Fermi constant, $\gamma_5$ is the Dirac matrix,
and $\rho_\mathrm{nuc}(r)$ is the {\em neutron} density distribution.
To be consistent with the previous calculations
the $\rho_\mathrm{nuc}(r)$  is taken to be a {\em proton} Fermi distribution employed in Ref.~\cite{BluJohSap90}.
The slight difference between the neutron and proton distributions will
be addressed in the conclusion of this section. The dominant contribution
to the PNC-amplitude, Eq.~(\ref{Eqn_E_PNC}), comes from intermediate states $6P_{1/2}$ and $7P_{1/2}$.
In Table~\ref{Tab_HwCorr} we present calculated third-order Breit
corrections to the relevant matrix elements of weak interaction.
Apparently, the dominant part of the Breit correction arises
from modifications at the Hartree-Fock level and in random-phase approximation (PRA).
All the corrections add coherently, and we do not expect that omitted higher-order
diagrams to be important. In fact, third-order RPA corrections are a few times smaller
than those in the second order, hinting at a good convergence of the present technique.

\begin{table}
\centering \caption{Breit corrections to matrix elements of weak interaction for $^{133}$Cs
in units of $10^{-11} i |e| a_0 (-Q_{\rm W}/N)$.
See caption of Table~\protect\ref{Tab_Acorr} for description
of columns.
\label{Tab_HwCorr}}
\begin{ruledtabular}
\begin{tabular}{lldddd}
\multicolumn{1}{c }{Transition}               &
\multicolumn{1}{c }{CHF}  %
& \multicolumn{1}{c }{$\delta (H_{\mathrm W})_{\rm HF}^{\rm I}$ } %
& \multicolumn{1}{c }{$\delta (H_{\mathrm W})_{\rm RPA}^{\rm II+III}$ } %
& \multicolumn{1}{c }{$\delta (H_{\mathrm W})_{\rm BO}^{\rm III}$ }%
& \multicolumn{1}{c }{$\delta (H_{\mathrm W})_{\rm Total}$} \\
\colrule %
$6S_{1/2}-6P_{1/2}$& 0.03159 &-0.00010   & -0.00015  &    -0.000029  &    -0.00028   \\
$6S_{1/2}-7P_{1/2}$& 0.01891 &-0.000058  & -0.000091 &   -0.000014  &    -0.00016   \\
$7S_{1/2}-6P_{1/2}$& 0.01656 &-0.000053  & -0.000081 &   -0.000013  &    -0.00015   \\
$7S_{1/2}-7P_{1/2}$& 0.00991 &-0.000031  & -0.000048 &   -0.0000061 &    -0.000085  \\
\end{tabular}
\end{ruledtabular}
\end{table}

The Breit correction to the PNC amplitude is determined by combining induced
modifications in matrix elements and energy denominators. The required
corrections are summarized in Table~\ref{Tab_bigPNC}. The tabulated
dipole amplitudes are related to the reduced matrix elements in Table~\ref{Tab_Dcorr}
as
\[
\langle nS_{1/2}| D | n'P_{1/2} \rangle =
 \langle nS_{1/2}|| D || n'P_{1/2} \rangle/\sqrt{6} \, .
\]

Before proceeding to the correlated calculations, it is worth examining the Breit contribution
to the PNC amplitude at the Hartree-Fock level.
Most of the Breit contribution to the PNC amplitude can be determined by limiting
the summation over intermediate states in Eq.~(\ref{Eqn_E_PNC})
to the two lowest valence $P_{1/2}$ states: $6P_{1/2}$ and $7P_{1/2}$.
In the CHF approximation
one then finds $E_\mathrm{PNC} =-0.6888$ (90\% of the total value).
The lowest-order corrections to matrix elements and energy denominators
calculated as differences between Breit-CHF and CHF values
are listed in Table~\ref{Tab_bigPNC}.
The resultant BCHF corrections to $E_\mathrm{PNC}$ are:
\begin{eqnarray}
E_\mathrm{PNC}^{\rm HF} (\delta H_{\mathrm W}) &=&  0.0022 \; ( 0.32\%) \, , \nonumber \\
E_\mathrm{PNC}^{\rm HF}  (\delta D) &=&    0.0020           \; ( 0.29\% ) \, , \label{Eqn_corr_HF}\\
E_\mathrm{PNC}^{\rm HF}  (\delta E) &=&   -0.0019           \; (-0.28\% )  \, . \nonumber
\end{eqnarray}
Here the percentage values in parentheses are taken with respect to
the DHF value of PNC amplitude.
The sum of these three terms leads to $\delta E_\mathrm{PNC} =0.0023$.
Inclusion of intermediate states beyond $6P_{1/2}$ and $7P_{1/2}$
leads to a small additional  modification  to $\delta E_\mathrm{PNC}$
of -0.00004. The obtained lowest-order result is
in agreement with the 0.002 correction found by
Blundell et al.~\cite{BluJohSap90}. In addition to the lowest-order
the Breit correction in Ref.~\cite{BluJohSap90} also
contained small random-phase-approximation diagrams in the Coulomb interaction
for matrix elements of the weak interaction. The two-body Breit interaction
has been disregarded in Ref.~\cite{BluJohSap90}. In the following discussion
we will include these omitted effects.
Note that if experimental energies
(which effectively include the Breit interaction) are used in the energy
denominators of Eq.~(\ref{Eqn_E_PNC}),
then the $E_\mathrm{PNC} (\delta E)$ term must be excluded and the total
correction becomes twice as large: $\delta E_\mathrm{PNC} =0.0042$.

\begin{table}
\caption{Breit corrections to electric-dipole amplitudes, weak interaction matrix elements,
and energy intervals; $\delta X, \rm{I} \equiv X_{\rm BCHF} - X_{\rm CHF}$, and
$\delta X, \rm{I+II+III}$ are the differences in the third order of MBPT.
Weak matrix elements are expressed in units of $10^{-11} i |e| a_0 (-Q_{\rm W}/N)$
and energies and dipole amplitudes in atomic units.
\label{Tab_bigPNC}}
\begin{ruledtabular}
\begin{tabular}{ldddd}
&
\multicolumn{1}{c}{$6S_{1/2}-6P_{1/2}$ } &
\multicolumn{1}{c}{$6S_{1/2}-7P_{1/2}$} &
\multicolumn{1}{c}{$7S_{1/2}-6P_{1/2}$} &
\multicolumn{1}{c}{$7S_{1/2}-7P_{1/2}$} \\
\hline
$ H_{\rm W}$, DHF       &   0.03159  &  0.01891 &  0.01656  &  0.009913 \\
$\delta H_{\rm W}$, I    & -0.00010  & -0.00006 & -0.00005  & -0.000031 \\
$\delta H_{\rm W}$,
I+II+III                &  -0.00028  & -0.00016 & -0.00015  & -0.000085  \\[3pt]
$ D $,      DHF          &   2.1546   &  0.15176 &  1.8017   &  4.4944   \\
$\delta D$, I            &   0.0001   &  0.00073 &  0.0019   & -0.0004   \\
$\delta D$, I+II+III     &  -0.0004   &  0.00077 &  0.0020   & -0.0012   \\[3pt]
$ \Delta E  $, DHF       &  -0.041752 & -0.085347&  0.030429 & -0.013166  \\
$ \delta \Delta E$, I    &  -0.000020 &  0.000003& -0.000030 & -0.000007  \\
$ \delta \Delta E$, I+II &  -0.000045 & -0.000023& -0.000034 & -0.000012  \\
\end{tabular}
\end{ruledtabular}
\end{table}

With further examination of the modifications of {\em individual} uncorrelated matrix
elements summarized in Table~\ref{Tab_bigPNC}, one notices the following. \\
(i) Weak interaction matrix elements are each reduced
in absolute value by 0.3\%, which is directly reflected in a 0.3\% correction to
the PNC amplitude.\\
(ii) Modification of dipole amplitudes is strongly
nonuniform. There are substantial corrections only to the $6S_{1/2}-7P_{1/2}$
(0.5\%) and
$7S_{1/2}-6P_{1/2}$ (0.1\%) matrix elements.
The large 0.5\% Breit correction to $\langle 6S_{1/2} | D | 7P_{1/2}  \rangle$
provides partial resolution to a long-standing 1.5\%
discrepancy of spectroscopic experiment~\cite{ShaMonKhl79} and {\em ab initio} Coulomb-correlated
calculations~\cite{DzuFlaKra89,BluJohSap91,SafJohDer99}.
 \\
(iii) The largest modification in the energy denominators is
0.1\% for $E_{7S}-E_{6P}$; however, this leads to a 0.3\% correction $E_\mathrm{PNC} (\delta E)$.
As recently emphasized  by~\citet{DzuFlaSus97},
such large sensitivity of the resulting PNC amplitude to small variations in individual
atomic properties entering Eq.~(\ref{Eqn_E_PNC}) arises due to a cancellation of relatively large
terms in the sum over states.

It is well known that correlations caused by residual Coulomb interactions not included in the
Hartree-Fock equations can lead to substantial modifications of the lowest-order
values. For example, the weak matrix element
$\langle 6S_{1/2} | H_{\rm W} | 6P_{1/2}  \rangle$
is increased by a factor of 1.8 by correlations due to residual Coulomb interactions.
As demonstrated in the previous sections, correlations are also  important for a
proper description of the Breit corrections.
Examination of the third-order corrections listed in
Table~\ref{Tab_bigPNC} reveals that the
corrections to weak interaction matrix elements become three times larger than those
in the lowest order. Using third-order matrix elements and second-order energies the
following {\em ab initio} corrections are determined:
\begin{eqnarray*}
E_\mathrm{PNC} (\delta H_{\mathrm W}) &=&  0.0043 \, ,\\
E_\mathrm{PNC} (\delta D) &=& 0.0035 \, ,\\
E_\mathrm{PNC} (\delta E) &=& -0.0028 \, .
\end{eqnarray*}
Thus the lowest-order corrections given in Eq.~(\ref{Eqn_corr_HF}) are amplified in
higher orders. The calculated Breit corrections to the PNC amplitude is expected
to be insensitive to the omitted higher-order diagrams. Such uncertainty
can arise if the calculated Breit corrections to leading classes of many-body diagrams
cancel.
Indeed, the calculated
Breit corrections always add coherently for matrix elements of weak interaction.
There are also no strong cancellations
between various Breit corrections to the relevant dipole amplitudes
$\langle nS_{1/2} | D | n'P_{1/2}  \rangle$ (see Table~\ref{Tab_Dcorr}).
Corrections to energy denominators are also stable with respect to the omitted
higher-order contributions. For example, in Section~\ref{Sec_En} we found
that the Breit correction to the energy of $7S_{1/2}$ valence  state due
to cancellations of calculated contributions is small (-0.26 cm$^{-1}$)
enhancing possible effect of smaller higher-order corrections. However, in
the calculation of the term $E_\mathrm{PNC} (\delta E)$
this (unstable) correction substantially appears only in a combination with
a 25 times larger and stable Breit correction (7.1 cm$^{-1}$) to the energy of $6P_{1/2}$ state.

We further improve the accuracy  of the calculation by combining all-order Coulomb-correlated
matrix elements and experimental energy denominators tabulated in Ref.~\cite{BluJohSap90}  with
the present third-order Breit corrections. The results are:
\begin{eqnarray}
E_\mathrm{PNC} (\delta H_{\mathrm W}) &=&  0.0047 \; (0.5\%) \, , \nonumber \\
E_\mathrm{PNC} (\delta D) &=&    0.0037          \;  (0.4\%) \, . \label{Eqn_corr_III} \\
E_\mathrm{PNC} (\delta E) &=&   -0.0030          \;  (-0.3\%) \, . \nonumber
\end{eqnarray}
Here the values in parentheses are defined relative to the Coulomb-correlated
PNC amplitude, Eq.~(\ref{Eqn_refVal}). The total
Breit correction to the PNC amplitude is
$\delta E_\mathrm{PNC}^{B}  = 0.0054$. This result was first reported in~\cite{Der00}.
Similar correction of $0.0053$
was obtained by~\citet{DzuHarJoh01}. \citet{KozPorTup01} found a
20\% smaller  correction of $0.004$; this discrepancy is most likely
due to the omission of the retarded part of the Breit interaction in
the calculations~\cite{KozPorTup01}.

If the experimental energies (which incorporate Breit corrections by definition)
are used in the energy denominators the term $E_\mathrm{PNC} (\delta E)$
should be excluded and the semiempirical Breit correction becomes
$\delta E_\mathrm{PNC}^\mathrm{B, s.e.}  = 0.0084$.
A minor difference of our treatment~\cite{Der00} of the
Breit correction to $E_\mathrm{PNC}$ and
Ref.~\cite{DzuHarJoh01,KozPorTup01} is the {\em interpretation}
of results of previous Coulomb-correlated calculations~\cite{BluJohSap90,DzuFlaSus89}.
Authors~\cite{DzuHarJoh01,KozPorTup01} assert that the {\em ab initio}
0.6\% correction ( 0.4\% in Ref.~\cite{KozPorTup01} ) should be used
to augment $E_\mathrm{PNC}^{C}$, Eq.~\ref{Eqn_refVal}. Our approach is
to exclude term $E_\mathrm{PNC} (\delta E)$. The difference in the
two {\em interpretations} arises due to the difficulty of accounting
for higher order Coulomb diagrams. Some semiempirical ``fitting'' or
``scaling'' procedure is used in practice to mimic the effect of
the omitted contributions. Since only energies are known with
a very high precision from experiments, experimental energies
play a central role in such analysis.  For example, the experimental
energies were employed in eight out of ten  test cases in the scatter
analysis in the Table IV of Ref.~\cite{BluJohSap90} (Phys.\ Rev.\ D)  based
on Eq.~(\ref{Eqn_E_PNC}).

The above discussion demonstrates some arbitrariness
encountered in the analysis of theoretical values and
assignment of theoretical uncertainty when a semiempirical adjustment
of {\em ab initio} values
is attempted. We notice that authors of
Ref.~\cite{DzuHarJoh01,KozPorTup01} argue that the theoretical uncertainty
of the PNC amplitude is in the order of 1\%, therefore a 0.3\% difference
between the two different interpretations of the Breit correction is
irrelevant at this level. We believe that the
most convincing error estimate would be a repetition of the scatter analysis
originally performed in Ref.~\cite{BluJohSap90} but with
a Breit correction to the involved quantities included.
Such calculation is beyond the scope of the present work.
However, we expect that the resulting uncertainty
would be better than the original 1\% assigned to the
results~\cite{BluJohSap90,DzuFlaSus89} because of the better
theory-experiment agreement for dipole amplitudes (see \cite{BenWie99})
and hyperfine-structure constants (see Table~\ref{Tab_AExpComp}).
Additional QED and especially neutron skin effects can further modify the
value of $E_\mathrm{PNC}$ in a way that cannot be mimicked by
the suggested analysis.

The parity violation in atoms is dominated by the Z-boson exchange between
atomic electrons and {\em neutrons}. However the reference value $E^C_\mathrm{PNC}$,
Eq.~(\ref{Eqn_refVal}), is
based on the empirically deduced {\em proton} distribution.
The difference between the proton and neutron distributions is visualized as
the neutron ``skin'' or ``halo''. Here we update our previous treatment
of the neutron skin correction with the most recent data from the literature.
This correction was estimated in Ref.~\cite{BluJohSap90} but was not
included in the final value for the PNC amplitude.
It can be shown that the neutron skin correction
$\delta E_\mathrm{PNC}^\mathrm{n.s.}$
does not
depend on the electronic structure, therefore it can be
parameterized as
\begin{equation}
   \frac{ \delta E_\mathrm{PNC}^\mathrm{n.s.} }
   { E_\mathrm{PNC} } \approx -\frac{3}{7}
   \left( \alpha Z \right)^2 \frac{\Delta R_{np}} { R_{p} } \, .
\end{equation}
Here $R_p$ is the root-mean-square (rms) radius of proton distribution
and $\Delta R_{np}$ is the difference between rms radii of neutron and
proton distributions. This expression can be easily derived from analysis
by \citet{ForPanWil90}.
From  {\em nonrelativistic}
nuclear-structure calculations  \citet{PolWel99}
concluded $\Delta R_{np}/R_{p} = 0.016$ or 0.022 depending on the
model of nuclear forces.
The calculations \cite{VreLalRin00,PanDas00} of nuclear distributions were
{\em relativistic} and the corrections as twice as large
$\Delta R_{np}/R_{p} = 0.043-0.053$ were found. Therefore nonrelativistic calculations
led to $\delta E_\mathrm{PNC}^\mathrm{n.s.}/E_\mathrm{PNC}$ of -0.1\% and
relativistic determinations to -0.3 -- -0.4\%. The latter values are comparable
to the experimental error bar  of the PNC
amplitude~\cite{WooBenCho97,BenWie99} and unfortunately it is difficult
to assess the accuracy of the nuclear-structure calculations.

Here we propose an alternative analysis of the neutron skin correction allowing
to estimate the error bar.
Indeed \citet{TrzJasLub01} very recently
deduced neutron density distributions from experiments with antiprotonic atoms and
concluded that
\begin{equation}
 \Delta R_{np} = (-0.04 \pm 0.03) + (1.01 \pm 0.15) \, \frac{N-Z}{A} \, \,
 \mathrm{fm} \, .
 \label{Eqn_Rnp}
\end{equation}
Here $N$ and $A$ are the neutron and the mass numbers.
Although experimental data for $^{133}$Cs do not enter the
analysis\cite{TrzJasLub01},
a wide range of stable nuclei was investigated.
Assuming that this relation holds for $^{133}$Cs we find $\Delta R_{np} = 0.13(4)$ fm.
For $^{133}$Cs $R_p=4.807$ fm~\cite{JohSof85} leading to
\begin{equation}
   \frac{ \delta E_\mathrm{PNC}^\mathrm{n.s.} }
   { E_\mathrm{PNC} } = -0.0019(8) \, . \label{Eqn_ns}
\end{equation}
Therefore the neutron skin corrects the PNC amplitude by -0.2\% with an
error bar of 30\%. This uncertainty contributes only 0.06\% to an
error budget of the observed weak charge, i.e. the proposed determination of
the neutron skin correction will be adequate until the  0.1\% level
of overall accuracy is reached.

Combining the calculated semiempirical 0.9\% Breit correction
with the reference Coulomb-correlated value, Eq.(\ref{Eqn_refVal}), and
the neutron skin correction, Eq.(\ref{Eqn_ns}),
one obtains the parity-nonconserving amplitude
\begin{equation}
 E_\mathrm{PNC}(^{133}{\rm Cs} ) = -0.8974 \times 10^{-11} i(-Q_{\rm W}/N) \, .
 \label{Eqn_EpncTh}
\end{equation}
In section~\ref{Sec_OffDiag} we concluded that the  result for the
off-diagonal hyperfine structure matrix element $M_{\rm hf}$
by \citet{DzuFla00} is
not affected by the Breit correction. Using their value of $M_{\rm hf}$
together with the experimental results~\cite{WooBenCho97,BenWie99} we arrive at
\begin{equation}
  E_\mathrm{PNC}(^{133}{\rm Cs} ) = -0.8354(33) \times 10^{-11} \, \mathrm{a.u.}
\label{Eqn_EpncExp}
\end{equation}
From the values above, the observed weak charge is
\[
Q_{\rm W}(^{133}{\rm Cs} ) = -72.61(28)_{\rm expt}(73)_{\rm theor} \, .
\]
This value differs from the prediction~\cite{MarRos90} of the
Standard Model $Q_{\rm W}^{\rm SM}=-73.20(13)$ by 0.7$\sigma$,
versus 2.5$\sigma$  of Ref.~\cite{BenWie99}, where $\sigma$ is calculated
by taking experimental and theoretical uncertainties in quadrature.
Here we assigned 1\% uncertainty to the theoretical PNC amplitude,
Eq.~(\ref{Eqn_EpncTh}).
The deviation stands at $1.3\sigma$ if 0.4\% theoretical uncertainty
is assumed as discussed by \citet{BenWie99}. Following Ref.~\cite{Der00},
similar conclusion has been reached in Ref.~\cite{DzuHarJoh01,KozPorTup01}.

\section{Conclusion}
In this work we presented a relativistic
many-body formalism for treating correction from the Breit
interaction. Numerical evaluation of the Breit corrections
to a number of properties of cesium atom were carried out. In particular
we considered energies, hyperfine-structure constants, electric-dipole
transition amplitudes and $6S-7S$ parity violating amplitude.
We demonstrated that the Breit corrections to correlations are
as important as the modifications at the lowest-order Dirac-Hartree-Fock level.
This work supplements Ref.~\cite{Der00} with additional numerical results.
The present treatment has been based on third-order relativistic
many-body perturbation theory.
In a few cases we observed intricate cancellations between the lowest-order and
higher-order corrections. These are the counterintuitive cases where the most advanced
methods of many-body perturbation
theory originally developed for residual Coulomb interaction will have to be employed
to obtain an adequate description of a small Breit correction.

Is it possible to test the accuracy of the theoretical treatment of the Breit contribution for alkali-metal
structure? One could consider the nonrelativistically forbidden magnetic-dipole
transitions $nS_{1/2} -n'S_{1/2}$. A second-order analysis~\cite{SavDerBer99} demonstrated exceptionally
large contributions from the Breit interaction and negative-energy states for such transitions;
more accurate all-order calculations  would be desirable. At the same time an accurate experimental
value for the $7S-6S$ transition in Cs is available~\cite{BenWie99}.

We determined Breit correction to parity non-conserving (PNC) amplitude of
the  $6S-7S$ transition. The calculated correction resolves most of the
discrepancy~\cite{BenWie99} between the standard model prediction
and atomic PNC determination of the $^{133}$Cs weak charge.

Breit correction to the PNC amplitude is one of the smaller contributions
which needed to be addressed in order to reach the next level of accuracy
in {\em ab initio} calculations required for refined interpretation
of parity violation. In this work we also evaluated and
constrained the neutron ``skin'' correction.
As discussed in Ref.~\cite{Sus01,DzuHarJoh01,KozPorTup01},
the remaining corrections which can contribute at a few 0.1\%
are due to higher-order many-body diagrams in the Coulomb interaction
and QED corrections.

\begin{acknowledgments}
 I would like to thank W. R. Johnson, V. A. Dzuba, and S. G. Porsev for discussions.
The developed numerical code was partially based on programs by Notre
Dame group led by W.R. Johnson.
This work was supported in part by the National Science Foundation and
by the Chemical Sciences, Geosciences and
Biosciences Division of the Office of Basic Energy Sciences,
Office of Science, U.S. Department of Energy.
\end{acknowledgments}


\end{document}